\documentclass[11pt]{article}

\usepackage{graphicx}
\usepackage{subcaption}
\usepackage{mwe}
\usepackage{xcolor}
\usepackage{longtable}
\usepackage{supertabular}
\usepackage{pspicture}
\usepackage{filecontents}
\usepackage{pst-bar}
\usepackage{bm,amsfonts,amssymb,amsmath,float}
\usepackage{sectsty}
\usepackage{colortbl}
\usepackage{multirow}
\usepackage{latexsym}
\usepackage{epsfig}
\usepackage{epstopdf}
\usepackage{threeparttable}
\usepackage{caption}
\usepackage{enumerate}
\usepackage{diagbox}
\usepackage[normalem]{ulem}
\usepackage{bm}
\usepackage{mathtools}
\usepackage{enumitem}
\usepackage{algorithm}
\usepackage{algpseudocode}
\usepackage{afterpage}
\usepackage{bbm}
\usepackage{algpseudocodex}
\usepackage{placeins}
\usepackage{makecell}
\usepackage{appendix}
\usepackage{indentfirst} 

\newenvironment{proof}{\noindent{\bf Proof:}}{\hfill\fbox{}\vspace*{1mm}}
\newtheorem{theorem}{Theorem}[section]
\newtheorem{example}[theorem]{Example}
\newtheorem{lemma}[theorem]{Lemma}
\newtheorem{definition}[theorem]{Definition}
\newtheorem{remark}[theorem]{Remark}
\newtheorem{proposition}[theorem]{Proposition}
\newtheorem{corollary}[theorem]{Corollary}
\newtheorem{problem}[theorem]{Problem}
\newtheorem{idea}[theorem]{Idea}

\makeatletter
\newcommand\nextitem[1]{%
  \setcounter{\@enumctr}{#1}%
  \addtocounter{\@enumctr}{-1}%
}
\makeatother

\usepackage{ulem} 
\usepackage{hyperref}
\usepackage[mathlines, displaymath]{lineno} 

\usepackage{graphics}
\usepackage{amstext}
\usepackage{amscd}
\usepackage{rotating}
\usepackage{makeidx}
\usepackage{MnSymbol}
\usepackage{mdwlist}
\usepackage{fancyvrb}
\usepackage{mdframed}

\usepackage{tikz}
\usetikzlibrary{arrows.meta}

\begin{document}
\renewcommand{\labelenumi}{\arabic{enumi}.}
\renewcommand{\labelenumii}{\arabic{enumi}.\arabic{enumii}}
\renewcommand{\labelenumiii}{\arabic{enumi}.\arabic{enumii}.\arabic{enumiii}}
\renewcommand{\labelenumiv}{\arabic{enumi}.\arabic{enumii}.\arabic{enumiii}.\arabic{enumiv}}

\title{On the Minimal Controllability of Boolean Networks With $k$-Regular Dependency Graphs}

\author{Christopher H. Fok$^1$, Chi-Wing Wong$^1$, Wai-Ki Ching$^1$$^*$}
\date{
   $^1$Department of Mathematics, The University of Hong Kong, Pokfulam Road, Hong Kong\\
   $^*$wching@hku.hk}

\maketitle

\begin{abstract}
Boolean networks (BNs) are widely used to model gene regulatory networks and other biological systems. In this paper, we investigate the minimal controllability problem for BNs subject to degree constraints. We first establish three new best-case upper bounds for XOR-BNs whose dependency graphs are $k$-regular. We then introduce classes of BNs that are invariant under cyclically partite transformation (CPT-invariant) and prove theorems connecting BN controllability with the surjectivity of vector-valued Boolean functions. Finally, by combining an analysis of this surjectivity with a probabilistic argument, we derive a new general lower bound on the minimum number of control nodes required for majority-type threshold BNs with $(2k+1)$-regular dependency graphs.

\end{abstract}

\section{Introduction}\label{section:intro}

Boolean networks (BNs) were introduced by Kauffman as models of gene regulatory networks \cite{KauffmanS.A.1969Msae}. Their applications have since expanded to other biological processes and systems, including T-cell receptor signaling \cite{Saez2007}, apoptosis \cite{apoptosis}, and the yeast cell-cycle network \cite{yeastrobust}. Beyond providing insights into the dynamics of biological systems, BN models offer a framework for developing intervention strategies that steer these systems from undesirable states toward desirable ones. These applications have motivated the development of the control theory for BNs, an area that has attracted growing research interest in recent years \cite{AkutsuTree, ControlAndObserve, CoarsestPartition, StuckAt, AsympObserve, MinRealization, ZhongAsynchronous}.

An important direction in the control theory of BNs is the minimal controllability problem, which seeks the smallest set of control nodes that renders a given BN controllable (see Problem \ref{problem:BN_min_control} in Section \ref{section:prelim}). This problem and related control questions have been investigated in several studies. Kim et al. developed a general algorithm for computing a control kernel, defined as a minimal set of nodes whose control drives the network to a desired state \cite{ControlKernel}. Focusing on attractors as target states, Hou et al. showed, under a reasonable assumption, that $O(\log_{2}N + \log_{2}M)$ driver nodes suffice to control a BN, where $N$ denotes the number of nodes and $M$ denotes the number of attractors \cite{HouAttractor}. Through an analysis of 49 biological network models, Borriello and Daniels provided numerical evidence that the average size of a control kernel scales logarithmically with the number of attractors \cite{Borriello}. For conjunctive BNs, Weiss et al. investigated minimal controllability and established a graph-theoretic characterization of controllability, together with a quadratic-time algorithm for testing it \cite{WeissConjunctive}. Zhu et al. introduced the concept of strong structural controllability for BNs and proved that the corresponding minimal controllability problem is NP-hard \cite{BNStrongStruct}. In another study, Zhu et al. developed an efficient procedure for identifying a minimal set of controlled nodes that ensures global stabilization \cite{ZhuStabilizable}. Sun et al. derived bounds for the minimal controllability problem in BNs involving the XOR operator, the $k$-ary AND function, the $k$-ary AND function with negation, and nested canalizing Boolean functions \cite{SunMinControl}. Fok et al. likewise established bounds for this problem in BNs involving Boolean threshold functions and the XOR operator \cite{FokMinControl}.

Building on the work of Fok et al. \cite{FokMinControl}, this paper establishes further results on the minimal controllability of majority-type threshold BNs and XOR-BNs. The study of threshold BN control is motivated by the ability of these networks to capture activation and inhibition effects in a range of biological systems \cite{yeastrobust, FissionYeast, PBTN, Arabidopsis, FlowerEvo}. Our results may therefore offer insights into the control of real biological systems. Beyond biological modeling, Boolean threshold functions also arise in theoretical models of artificial neural networks \cite{AnthonyNeural, SiuNeural}. Our results for threshold BNs focus exclusively on majority-type Boolean functions, which have important applications in social network modeling and digital circuit design \cite{MajorityVote, MajorityLogic, TieBreak}. Our results may thus also be relevant to research in those areas. In contrast, our investigation of the minimal controllability of XOR-BNs is motivated by theoretical interest.

Our contributions in this paper are threefold. Firstly, we derive three new best-case upper bounds for XOR-BNs with $k$-regular dependency graphs. Secondly, we define the notion of BN sets which are invariant under cyclically partite transformation (CPT-invariant), and proved theorems which relate the controllability of BNs and the surjectivity of vector-valued Boolean functions. Thirdly, by considering the surjectivity of vector-valued Boolean functions and using a probabilistic argument, we prove a new general lower bound for the minimal number of control nodes of majority-type threshold BNs with $(2k+1)$-regular dependency graphs.

The rest of this paper is organized as follows. Section \ref{section:prelim} introduces the necessary preliminary concept studied in this paper. Section \ref{section:XOR} provides the three best-case upper bounds for XOR-BNs with $k$-regular dependency graphs. Section \ref{section:CPT_invariant} proposes the concept of CPT-invariant sets of BNs and proves theorems that relate the sizes of control-node sets of BNs and the surjectivity of vector-valued Boolean functions. Section \ref{section:MAJORITY} derives a general lower bound for majority-type threshold BNs with $(2k+1)$-regular dependency graphs. In Section \ref{section:conclusions}, we provide our concluding remarks and propose some possible future directions of research.

\section{Preliminaries}\label{section:prelim}

A Boolean network (BN) is a discrete-time dynamical system which consists of a finite number of nodes $x_{1}$, $x_{2}$, \ldots, $x_{n}$ and a finite number of Boolean functions $f_{1}, f_{2}, \ldots, f_{n} : \{ 0, 1 \}^{n} \to \{ 0, 1 \}$. For all non-negative integers $t$, we denote the state of a BN at time $t$ by $\mathbf{x}(t) \coloneqq (x_{1}(t), x_{2}(t), \ldots, x_{n}(t))$. When there is no control signals acting on the BN, the evolution of a BN follows the rules given below:
\begin{equation}
x_{i}(t + 1) = f_{i}(\mathbf{x}(t)) \quad \text{for $i = 1, 2, \ldots, n$.}
\end{equation}
We let $F \coloneqq (f_{1}, f_{2}, \ldots, f_{n}) : \{ 0, 1 \}^{n} \to \{ 0, 1 \}^{n}$. 

Let $U \subseteq \{ x_{1}, x_{2}, \ldots, x_{n} \}$ be the set of control nodes of a BN. When we exert control signals on a BN, the evolution of the BN can be described below:
\begin{align}
x_{i}(t + 1) &= f_{i}(\mathbf{x}(t)),			& &\forall x_{i} \notin U, \\
x_{i}(t + 1) &= u_{i}(t) \oplus f_{i}(\mathbf{x}(t)),	& &\forall x_{i} \in U,
\end{align}
where $\oplus$ is the XOR logical operator and for all non-negative integers $t$, the control signals $u_{i}(t)$ can be either $0$ or $1$.

Now, we define some important concepts.
\begin{definition}\label{def:relevant_input}
Let $\varphi : \{ 0, 1 \}^{k} \to \{ 0, 1 \}$ be a Boolean function. The $i$-th input of $\varphi$ is said to be relevant if there exist $a_{1}, \ldots, a_{i-1}, a_{i+1}, \ldots, a_{k} \in \{ 0, 1 \}$ such that 
$\varphi(\mathbf{v}, 0, \mathbf{u}) \neq \varphi(\mathbf{v}, 1, \mathbf{u})$, 
where $\mathbf{v} \coloneqq (a_{1}, \ldots, a_{i-1})$,
$\mathbf{u} \coloneqq (a_{i+1}, \ldots, a_{k})$.
\end{definition}

\begin{definition}
The dependency graph of a BN is defined to be the directed graph $G = (V, E)$ such that 
$V \coloneq \{ x_{1}, x_{2}, \ldots, x_{n} \}$ and for all $x_{i}, x_{j} \in V$, there exists an edge from $x_{i}$ to $x_{j}$ 
iff
$x_{i}$ is a relevant input of the Boolean function $f_{j}$.
We require that each node of $G$ has at most one self-loop, and that for all $x_{i} \neq x_{j}$, there is at most one edge from $x_{i}$ to $x_{j}$ in $G$.
\end{definition}

\begin{definition}
If an $n$-node BN with control-node set $U$ satisfies that for all $\mathbf{a}, \mathbf{b} \in \{ 0, 1\}^{n}$, there exist control signals which drive the BN from $\mathbf{x}(0) = \mathbf{a}$ to $\mathbf{x}(T) = \mathbf{b}$ for some non-negative integer $T$, then the BN is said to be controllable.
\end{definition}

\begin{problem}[Minimal Controllability Problem of BNs]\label{problem:BN_min_control}
Given a BN, find the smallest control-node set which makes the BN controllable.
\end{problem}

We also consider two concepts related to the sizes of the minimal control-node sets \cite{SunMinControl, SunObserve} (shown in Table \ref{table:meaning_of_bounds}).
\begin{table}[h!]
\centering
\caption{The meaning of general lower bounds and best-case  upper bounds.}\label{table:meaning_of_bounds}
\begin{tabular}{|l|l|}
\hline
\makecell[l]{General \\ lower bound} & \makecell[l]{For any BN satisfying certain properties, \\ the size of the minimal control-node set is at least $b$.} \\ \hline
\makecell[l]{Best-case \\ upper bound} & \makecell[l]{For some of the BNs satisfying certain properties, \\ the size of the minimal control-node set is at most $b$, \\ which is currently the smallest.} \\ \hline
\end{tabular}
\end{table}

\FloatBarrier

\section{Three New Best-Case Upper Bounds for XOR-BNs}\label{section:XOR}

In this section, we study the minimal controllability problem of XOR-BNs, which is defined below. 

\begin{definition}\label{def:XOR-BNs}
An $n$-node BN is said to be an XOR-BN if for all nodes $x_{i}$, there exist distinct nodes $x_{i_{1}}, x_{i_{2}}, \ldots, x_{i_{l}} \in \{ x_{1}, x_{2}, \ldots, x_{n} \}$ such that
\begin{equation}\label{eq:XOR-BN_update_rules}
x_{i}(t + 1) = x_{i_{1}}(t) \oplus x_{i_{2}}(t) \oplus \ldots \oplus x_{i_{l}}(t).
\end{equation}
\end{definition}

To prove the new best-case upper bounds for XOR-BNs, we will need some useful ideas and results in \cite{FokMinControl}. They are presented in Idea \ref{idea:new_perspective}, Theorem \ref{thm:Wu_spans_Fn_2} and Lemma \ref{lemma:M1}.

\begin{idea}\label{idea:new_perspective}
Since $\oplus$ is the addition operation of the finite field $F_{2}$, the functional dependencies of an $n$-node XOR-BN can be represented by a linear map from $F^{n}_{2}$ to $F^{n}_{2}$ or an $n \times n$ matrix over $F_{2}$.
For example, consider the $3$-node XOR-BN defined as
\begin{align}
x_{1}(t + 1) &= x_{2}(t) \oplus x_{3}(t)  \label{eq:XOR_example_1}\\
x_{2}(t + 1) &= x_{1}(t) \oplus x_{2}(t) \oplus x_{3}(t)  \label{eq:XOR_example_2} \\
x_{3}(t + 1) &= x_{1}(t) \oplus x_{3}(t)  \label{eq:XOR_example_3}
\end{align}
Then, the matrix $A \in F^{3 \times 3}_{2}$ associated with this XOR-BN is
\begin{equation}
A =
\left(
\begin{array}{ccc}
0 & 1 & 1 \\
1 & 1 & 1 \\
1 & 0 & 1
\end{array}
\right).
\end{equation}
Note that (\ref{eq:XOR_example_1}), (\ref{eq:XOR_example_2}) and (\ref{eq:XOR_example_3}) can also be expressed in the following way:
\begin{equation}
\mathbf{x}(t + 1) = A\mathbf{x}(t).
\end{equation}
\end{idea}

\begin{theorem}\label{thm:Wu_spans_Fn_2}
Let $\mathcal{B}$ be an $n$-node XOR-BN with $A \in F^{n \times n}_{2}$ as the associated matrix.
For all subset $U$ of nodes, $U$ renders $\mathcal{B}$ controllable iff the set
\begin{equation}
W_{U} \coloneqq 
\text{Span}\left( \bigcup_{i\, \text{s.t.}\, x_{i} \in U} \{ A^{k}\mathbf{e}_{i} : k \in \mathbb{Z}_{\geq 0} \} \right)
\end{equation}
equals $F^{n}_{2}$.
Here, $\mathbf{e}_{1}, \ldots, \mathbf{e}_{n} \in F^{n}_{2}$ denote the standard unit vectors.
\end{theorem}

\begin{lemma}\label{lemma:M1}
Let $\mathbf{v} \in F^{n}_{2} \setminus \{ \mathbf{0}_{n} \}$ and $A \in F^{n \times n}_{2}$.
There exists $k \geq 0$ such that $\mathbf{v}$, $A\mathbf{v}$, \ldots, $A^{k}\mathbf{v}$ form a basis of 
$\text{Span}(\left\{ A^{l}\mathbf{v} : l \in \mathbb{Z}_{\geq 0} \right\})$.
\end{lemma}

\begin{definition}\label{def:k-k-XOR-BNs}
An XOR-BN whose dependency graph is $k$-regular for some positive integer $k$ is called a $k$-$k$-XOR-BN. Note that in the matrix associated with a $k$-$k$-XOR-BN, every row has exactly $k$ 1-entries, and the same is true for every column.
\end{definition}

In \cite{FokMinControl}, Fok et al. proved the following two best-case upper bounds for $k$-$k$-XOR-BNs (Theorems \ref{thm:old_2-2-XOR-BNs} and \ref{thm:old_powers_of_2}). 

\begin{theorem}\label{thm:old_2-2-XOR-BNs}
For all integers $n \geq 3$, there exists a controllable $n$-node 2-2-XOR-BN with 1 control node.
\end{theorem}

\begin{theorem}\label{thm:old_powers_of_2}
For all integers $m \geq 2$ and $k \in [3, 2^{m} - 1]$, there exists a controllable $2^{m}$-node $k$-$k$-XOR-BN with 1 control node. 
\end{theorem}

In a moment, we will prove three more best-case upper bounds (Theorem \ref{thm:trivial_1_control_node} and Corollaries \ref{coroll:coprime_1_control_node_thm}, \ref{coroll:4-4-XOR-BNs}) which complement these previous results. To do so, we need to provide three definitions and one proposition.

\begin{definition}
We use $F_{2}[t]$ to denote the polynomial ring over $F_{2}$.
\end{definition}

\begin{definition}
Fix arbitrary $A \in F^{n \times n}_{2}$. Then, $F_{2}[A] \coloneqq \{ p(A) : p(t) \in F_{2}[t] \}$.
We remark that the identity matrix $I \in F^{n \times n}_{2}$ is the identity element in $F_{2}[A]$.
\end{definition}

\begin{definition}
Let $P \in F^{n \times n}_{2}$ such that $P\mathbf{e}_{n} = \mathbf{e}_{1}$ and for $i = 1, 2, \ldots, n-1$,
$P\mathbf{e}_{i} = \mathbf{e}_{i+1}$, where $\mathbf{e}_{1}, \mathbf{e}_{2}, \ldots, \mathbf{e}_{n}$ denote the standard unit vectors in $F^{n}_{2}$.
\end{definition}

\begin{proposition}\label{prop:P}
Fix arbitrary $A \in F^{n \times n}_{2}$. If $P \in F_{2}[A]$, then
$W \coloneqq \text{Span} \left( \left\{ A^{k}\mathbf{e}_{1} : k \in \mathbb{Z}_{\geq 0} \right\} \right) = F^{n}_{2}$.
\end{proposition}

\begin{proof}
Clearly, $\mathbf{e}_{1} \in W$. We will prove this result by showing that $\mathbf{e}_{2}, \mathbf{e}_{3}, \ldots, \mathbf{e}_{n} \in W$.

Fix arbitrary $m \in \{ 2, 3, \ldots, n \}$. Note that $P = \sum^{L}_{i=1} A^{k_{i}}$ where 
$k_{1}, k_{2}, \ldots, k_{L} \in \mathbb{Z}_{\geq 0}$. Then,
\begin{equation}
\mathbf{e}_{m} = P^{m-1}\mathbf{e}_{1} = \left( \sum^{L}_{i=1} A^{k_{i}} \right)^{m-1} \mathbf{e}_{1}
= \left( \sum^{M}_{j=1} A^{l_{j}} \right) \mathbf{e}_{1}
\end{equation}
for some $l_{1}, l_{2}, \ldots, l_{M} \in \mathbb{Z}_{\geq 0}$. 
Clearly, $\sum^{M}_{j=1} A^{l_{j}} \mathbf{e}_{1} \in W$. Hence, $\mathbf{e}_{m} \in W$. The proof is complete.
\end{proof}

\begin{theorem}\label{thm:trivial_1_control_node}
For all odd integers $n \geq 3$, there exists an $n$-node controllable $(n-1)$-$(n-1)$-XOR-BN with only one control node.
\end{theorem}

\begin{proof}
We will prove that there exists a matrix $A \in F^{n \times n}_{2}$ such that each row/column of $A$ has exactly 
$(n-1)$ 1-entries and $\mathbf{e}_{1}$, $A\mathbf{e}_{1}$, \ldots, $A^{n-1}\mathbf{e}_{1}$ are linearly independent. This theorem will then follows directly from Theorem \ref{thm:Wu_spans_Fn_2}.

Let $J \in F^{n \times n}_{2}$ be the all-ones matrix.
Define $A = J + P$. Note that for $i = 1, 2, \ldots, n-1$, $A\mathbf{e}_{i} = \mathbf{1}_{n} + \mathbf{e}_{i+1}$.
Since $A\mathbf{1}_{n} = \mathbf{0}_{n}$, we have $A^{k}\mathbf{e}_{1} = \mathbf{1}_{n} + \mathbf{e}_{k+1}$ for 
$k = 1, 2, \ldots, n-1$. We will now show that $\mathbf{e}_{1}$, $\mathbf{1}_{n} + \mathbf{e}_{2}$, $\mathbf{1}_{n} + \mathbf{e}_{3}$, \ldots, $\mathbf{1}_{n} + \mathbf{e}_{n}$ are linearly independent. Suppose that
$a_{1}, a_{2}, \ldots, a_{n} \in F_{2}$ satisfy
\begin{equation}
a_{1}\mathbf{e}_{1} + \sum^{n}_{i = 2} a_{i}(\mathbf{1}_{n} + \mathbf{e}_{i}) = \mathbf{0}_{n}.
\end{equation}
Let $s \coloneqq \sum^{n}_{i = 2} a_{i}$. Coordinates 2, 3, \ldots, $n$ gives $a_{2} = a_{3} = \cdots = a_{n} = s$. Therefore,
\begin{equation}
\mathbf{0}_{n}
= a_{1}\mathbf{e}_{1} + \sum^{n}_{i = 2} a_{i} \mathbf{1}_{n} + \sum^{n}_{i=2} a_{i} \mathbf{e}_{i}
= a_{1}\mathbf{e}_{1} + s \sum^{n}_{i = 2} \mathbf{1}_{n} + \sum^{n}_{i=2} a_{i} \mathbf{e}_{i}
= \sum^{n}_{i=1} a_{i} \mathbf{e}_{i}.
\end{equation}
Since $\mathbf{e}_{1}$, $\mathbf{e}_{2}$, \ldots, $\mathbf{e}_{n}$ are linearly independent,  we have 
$a_{1} = a_{2} = \cdots = a_{n} = 0$.
Hence, $\mathbf{e}_{1}$, $A\mathbf{e}_{1}$, \ldots, $A^{n-1}\mathbf{e}_{1}$ are linearly independent.
The proof is complete.
\end{proof}

\begin{theorem}\label{thm:special_matrix}
Let $n> 3$ be an integer such that $\gcd(3, n) = 1$. There exists $A \in F^{n \times n}_{2}$ such that each row/column of $A$ has exactly three 1-entries and $\mathbf{e}_{1}$, $A\mathbf{e}_{1}$, \ldots, 
$A^{n-1}\mathbf{e}_{1}$ form a basis of $F^{n}_{2}$.
\end{theorem}

\begin{proof}
Consider the permutation matrix $S \in F^{n \times n}_{2}$ such that $S\mathbf{e}_{1} = \mathbf{e}_{2}$, $S\mathbf{e}_{2} = \mathbf{e}_{1}$ and for all $i \in [3, n]$, $S\mathbf{e}_{i} = \mathbf{e}_{i}$. Set
\begin{align}
T &\coloneqq I_{n} + P + P^{-1}, \label{eq:matrix_T} \\
A &\coloneqq TS. \label{eq:matrix_A}
\end{align}
Suppose that $\mathbf{x} \in F^{n}_{2}$ satisfies $T\mathbf{x} = \mathbf{0}_{n}$. With subscripts interpreted modulo $n$, we have
\begin{equation}\label{eq:m_terms_of_x}
x_{i} + x_{i-1} + x_{i+1} = 0
\end{equation}
for $i = 1, 2, \ldots, n$.
This implies that $x_{i+3} = x_{i}$. Since $\gcd(n, 3) = 1$, all coordinates of $\mathbf{x}$ are equal.
Substituting their common value $c$ into Eq. (\ref{eq:m_terms_of_x}), we get $3c = c = 0$.
Therefore, $T$ and hence $A = TS$ are invertible. This argument works over any field extension of $F_{2}$.

From now on, we will work over an algebraic closure $K$ of $F_{2}$. Assume for a contradiction that
$\mathbf{e}_{1}$, $A\mathbf{e}_{1}$, \ldots, $A^{n-1}\mathbf{e}_{1}$ are linearly dependent. Then, their span $W$ is a proper $A$-invariant subspace of $K^{n}$. Because $\text{dim}(W) + \text{dim}(W^{\perp}) = n$, $W^{\perp}$ is a non-zero $A^{T}$-invariant subspace of $K^{n}$. Since $K$ is algebraically closed, $A^{T}$ has an eigenvector 
$\mathbf{v} \in W^{\perp} \setminus \{ \mathbf{0}_{n} \}$. Let $A^{T}\mathbf{v} = \lambda \mathbf{v}$ 
where $\lambda \in K$. Because $\mathbf{e}_{1} \in W$, we have $v_{1} = 0$. Moreover, the invertibility of $A^{T}$ implies that $\lambda \neq 0$. Hence,
\begin{align}
&(TS)^{T}\mathbf{v} = \lambda\mathbf{v} \\
\Rightarrow &S^{T} T^{T}\mathbf{v} = \lambda\mathbf{v} \\
\Rightarrow &ST\mathbf{v} = \lambda\mathbf{v} \\
\Rightarrow &T\mathbf{v} = \lambda S\mathbf{v}. \label{eq:XOR_star_1}
\end{align}
Let $\sigma \coloneqq \sum^{n}_{i = 1} v_{i}$. Summing the coordinates on both sides of Eq. (\ref{eq:XOR_star_1}), we have
\begin{align}
&\sigma = \lambda \sigma \\
\Rightarrow &(1 + \lambda)\sigma = 0 \\
\Rightarrow &(1 + \lambda)\sigma^{2} = 0 \label{eq:XOR_star_2}
\end{align}

Note that
\begin{align}
\mathbf{v}^{T} T \mathbf{v}
&= \mathbf{v}^{T} (I_{n} + P + P^{-1}) \mathbf{v} \\
&= \sigma^{2} 
+ \sum^{n}_{k = 1} v_{k} v_{k+1}
+ \sum^{n}_{k = 1} v_{k} v_{k-1} \\
&= \sigma^{2}. \label{eq:XOR_star_3}
\end{align}
Because $v_{1} = 0$, we have
\begin{equation}\label{eq:XOR_star_4}
\mathbf{v}^{T}S\mathbf{v} = \sum^{n}_{i = 3} v^{2}_{i} = \sigma^{2} + v^{2}_{2}.
\end{equation}
Multiplying Eq. (\ref{eq:XOR_star_1}) on the left by $\mathbf{v}^{T}$, we get
\begin{align}
&\mathbf{v}^{T} T \mathbf{v} = \lambda \mathbf{v}^{T}S\mathbf{v}	& 	& \\
\Rightarrow&\sigma^{2} = \lambda(\sigma^{2} + v^{2}_{2})			& 	& \text{(by Eq. (\ref{eq:XOR_star_4}))} \\
\Rightarrow&(1 + \lambda)\sigma^{2} = \lambda v^{2}_{2}			&	& \\
\Rightarrow&0 = \lambda v^{2}_{2}.							& 	& \text{(by Eq. (\ref{eq:XOR_star_2}))}
\end{align}
Since $\lambda \neq 0$, $v_{2} = 0$. Therefore, $S\mathbf{v} = \mathbf{v}$. Then, Eq. (\ref{eq:XOR_star_1}) reduces to $T\mathbf{v} = \lambda\mathbf{v}$. Its second coordinate gives $v_{1} + v_{2} + v_{3} = \lambda v_{2}$, 
so $v_{3}= 0$. By successively applying the remaining coordinate equations, we get 
$v_{1} = v_{2} = \cdots = v_{n} = 0$. This contradicts with the fact that $\mathbf{v} \neq \mathbf{0}_{n}$. Thus, we conclude that $\mathbf{e}_{1}$, $A\mathbf{e}_{1}$, \ldots, $A^{n-1}\mathbf{e}_{1}$ are linearly independent over $K$ and hence $F_{2}$.
\end{proof}

\begin{corollary}\label{coroll:coprime_1_control_node_thm}
Let $n> 3$ be an integer such that $\gcd(3, n) = 1$. There exists a controllable $n$-node 3-3-XOR-BN with only 1 control node.
\end{corollary}

\begin{proof}
This result follows directly from Theorems \ref{thm:Wu_spans_Fn_2} and \ref{thm:special_matrix}.
\end{proof}

\begin{theorem}\label{thm:4-4-XOR-BNs}
Let $n > 4$ be an integer. There exists $A \in F^{n \times n}_{2}$ such that each row/column of $A$ has exactly four 1-entries and $\text{Span}(\{ A^{k}\mathbf{e}_{1} : k \in \mathbb{Z}_{\geq 0} \}) = F^{n}_{2}$.
\end{theorem}

\begin{proof}
By Proposition \ref{prop:P}, it suffices to show that for all $n > 4$, there exists $A \in F^{n \times n}_{2}$ such that each row/column of $A$ has exactly four 1-entries and $P \in F_{2}[A]$. We will split into four cases to consider.

Case 1: $n \equiv 0 \pmod 4$. Then, there exists an even integer $r \geq 3$ such that $n = 2r$. Define
\begin{equation}
A \coloneqq I + P + P^{2} + P^{r+2}.
\end{equation}
Note that 
\begin{equation}
A^{2}+I
= I + P^{2} + P^{4} + P^{2r+4} + I
= P^{2} + P^{4} + P^{4}
= P^{2}.
\end{equation}
Thus, $P^{2} \in F_{2}[A]$. Because $r + 2$ is even, we also have
\begin{equation}
P^{r+2} = (P^{2})^{\frac{r+2}{2}} \in F_{2}[A].
\end{equation}
Hence,
\begin{equation}
P = A + I + P^{2} + P^{r+2} \in F_{2}[A].
\end{equation}

Case 2: $n \equiv 2 \pmod 4$. Then, there exists an odd integer $r \geq 3$ such that $n = 2r$. Define
\begin{equation}
A \coloneqq I + P + P^{2} + P^{r+1}.
\end{equation} 
Note that
\begin{equation}
A^{2} + I
= I + P^{2} + P^{4} + P^{2r+2} + I
= P^{2} + P^{4} + P^{2}
= P^{4}.
\end{equation}
Hence, $P^{4} \in F_{2}[A]$, and
\begin{align}
(P^{4})^{\frac{r+1}{2}}	&=P^{2r+2} = P^{2} \in F_{2}[A], \\
P^{r+1}			&= (P^{2})^{\frac{r+1}{2}} \in F_{2}[A].
\end{align}
Therefore,
\begin{equation}
P = A + I + P^{2} + P^{r+1} \in F_{2}[A].
\end{equation}

Case 3: $n$ is odd and the multiplicative order of 2 modulo $n$ (denoted by $m$) is odd. Since $n\ |\ (2^{m} - 1)$, we have $P^{2^{m}} = P$. Set
\begin{equation}
A = I + P + P^{2} + P^{3} = (I + P)^{3}.
\end{equation}
Let $X \coloneqq I + P$ and $q \coloneqq 2^{m}$. Then,
\begin{equation}
X^{q}
= I + P^{q}
= I + P
= X.
\end{equation}
Since $m$ is odd, $q \equiv 2 \pmod 3$, so $d \coloneqq \frac{2q-1}{3}$ is a positive integer. Consequently,
\begin{equation}
A^{d}
= X^{3d}
= X^{2q-1}
= X^{q-1}X^{q}
= X^{q}
= X.
\end{equation}
Thus, $P = I + A^{d} \in F_{2}[A]$.

Case 4: $n$ is odd and the multiplicative order of 2 modulo $n$ (denoted by $m$) is even. Write $m = 2k$. Note that $k \geq 2$; otherwise, $m = 2$ and then 
\begin{align}
&1 \equiv 2^{m} \equiv 4 \pmod n \\
\Rightarrow\ &0 \equiv 3 \pmod n \\
\Rightarrow\ &n\ |\ 3,
\end{align}
but this is not possible because $n > 4$. Define
\begin{equation}
A \coloneqq I + P + P^{2} + P^{2^{k}}.
\end{equation}
Note that the residues $0, 1, 2, 2^{k}$ modulo $n$ are distinct. Indeed, $2^{k} \not\equiv 0 \pmod n$ because $n$ is odd. Moreover, if $2^{k} \equiv 1 \pmod n$, then $m\ |\ k$ which is not possible. Similarly, if $2^{k} \equiv 2 \pmod n$, then $2^{k-1} \equiv 1 \pmod n$ and $m\ |\ (k-1)$, which is again not possible. Therefore, we deduce that each row/column of $A$ has exactly four 1-entries.
Note that $P^{2^{2k}} = P$, which implies
\begin{align}
&I + A + A^{2} + A^{2^{2k}} \\
=& (P + P^{2} + P^{2^{k}}) + (I + P^{2} + P^{4} + P^{2^{k+1}}) + (I + P^{2^{k}} + P^{2^{k+1}} + P^{2^{2k}}) \\
=& P^{4} \in F_{2}[A].
\end{align}
Hence,
\begin{equation}
P = (P^{4})^{2^{m-2}} \in F_{2}[A]
\end{equation}
because $P^{2^{m}} = P$.
\end{proof}

\begin{corollary}\label{coroll:4-4-XOR-BNs}
For all integers $n > 4$, there exists an $n$-node controllable 4-4-XOR-BN with exactly one control node.
\end{corollary}

\begin{proof}
This result follows directly from Theorems \ref{thm:Wu_spans_Fn_2} and \ref{thm:4-4-XOR-BNs}.
\end{proof}

\section{Invariance Under Cyclically Partite Transformation}\label{section:CPT_invariant}

In this section, we derive new results about the relationship between the minimal controllability of BNs and the surjectivity of vector-valued Boolean functions (Theorem \ref{thm:restriction_becomes_surjective}, Definition \ref{def:CPT_invariant} and Theorem \ref{thm:surjectivity_and_min_control}).
Before we provide these original results, we present some previous results (Definition \ref{def:cyclically_s_partite_BNs} and Theorem \ref{thm:control_of_CSP_BNs}), which were proposed in an unpublished version of \cite{FokMinControl}.

\begin{definition}\label{def:cyclically_s_partite_BNs}
Let $s \geq 2$ be an integer. 
A BN is said to be \textit{cyclically $s$-partite} if its set of nodes can be partitioned into non-empty sets 
$V_{0}$, $V_{1}$, \ldots, $V_{s-1}$ such that in the dependency graph of the BN, each directed edge goes from a node in $V_{i}$ to a node in $V_{i+1\ (\text{mod}\ s)}$. 
$V_{i}$ is called the $i$-th layer of the BN. 
We denote the nodes in $V_{i}$ by $x^{i}_{1}$, $x^{i}_{2}$, \ldots, $x^{i}_{n_{i}}$. 
Let $j \coloneqq i-1\ (\text{mod}\ s)$. Note that 
$(x^{i}_{1}(t+1), \ldots, x^{i}_{n_{i}}(t+1)) = F_{i}(x^{j}_{1}(t), \ldots, x^{j}_{n_{j}}(t))$ for some 
$F_{i} : \{ 0, 1 \}^{n_{j}} \to \{ 0, 1 \}^{n_{i}}$.
We denote the BN by $\text{CP}(F_{0}, F_{1}, \ldots, F_{s-1})$.
\end{definition}

\begin{theorem}\label{thm:control_of_CSP_BNs}
Consider the BN $\text{CP}(F_{0}, F_{1}, \ldots, F_{s-1})$. 
For $i = 1, \ldots, s-1$, let $C_{i}$, $R_{i}$ be disjoint sets such that $C_{i} \cup R_{i} = V_{i}$. 
For all $\mathbf{x} \in \{ 0, 1 \}^{n_{i}}$, let $\mathbf{x}_{R_{i}}$ be the vector containing the entries of $\mathbf{x}$ corresponding to the nodes in $R_{i}$. 
If for $i = 1, \ldots, s-1$, $\{ \mathbf{y}_{R_{i}} : \mathbf{y} \in \text{Im}(F_{i}) \} = \{ 0, 1 \}^{\left| R_{i} \right|}$, 
then the control-node set $V_{0} \cup C_{1} \cup C_{2} \cup \ldots \cup C_{s-1}$ renders 
$\text{CP}(F_{0}, F_{1}, \ldots, F_{s-1})$ controllable.
\end{theorem}

\begin{theorem}\label{thm:restriction_becomes_surjective}
Consider an $n$-node BN with nodes $V = \{ x_{1}, x_{2}, \ldots, x_{n} \}$ whose update rule is given by
$F : \{ 0, 1 \}^{n} \to \{ 0, 1 \}^{n}$. For all subsets of nodes $U \subseteq V$, if $U$ renders the BN controllable, then the restriction $F_{V \setminus U} : \{ 0, 1 \}^{n} \to \{ 0, 1 \}^{n - |U|}$ of $F$ to nodes in $V \setminus U$ is a surjective function.
\end{theorem}

\begin{proof}
Without loss of generality, we can assume that $V \setminus U = \{ x_{1}, x_{2}, \ldots, x_{q} \}$ and
$U = \{ x_{q+1}, x_{q+2}, \ldots, x_{n} \}$. Fix arbitrary $\mathbf{u} \in \{ 0, 1 \}^{q}$ and choose any
$\mathbf{y} \in \{ 0, 1 \}^{n} \setminus \{ (\mathbf{u}, \mathbf{0}_{n-q}) \}$. Note that there exist control signals which drive the BN from $\mathbf{x}(0) = \mathbf{y}$ to $\mathbf{x}(T) = (\mathbf{u}, \mathbf{0}_{n-q})$, where 
$T \geq 1$. Let $\mathbf{x}(0), \mathbf{x}(1), \ldots, \mathbf{x}(T)$ be the states of the BN which occur during the control process. It is easy to see that $F_{V \setminus U}(\mathbf{x}(T-1)) = \mathbf{u}$. Since $\mathbf{u}$ is arbitrary, we deduce that $F_{V \setminus U}$ is surjective.
\end{proof}

\begin{definition}\label{def:CPT_invariant}
Let $\Pi$ be a set of BNs. If for each BN (whose update rule is denoted by $F : \{ 0, 1 \}^{n} \to \{ 0, 1 \}^{n}$) in 
$\Pi$, for all integers $s \geq 2$, the BN $\text{CP}(\underbrace{F, F, \ldots, F}_{s\ \text{times}})$ also belongs to 
$\Pi$, then $\Pi$ is said to be invariant under the cyclically partite transformation (CPT-invariant).
\end{definition}

\begin{example}
Let $f : \{ 0, 1 \}^{k} \to \{ 0, 1 \}$ be a Boolean function whose inputs are all relevant. The set of all BNs $B$ which satisfy that each node of $B$ depends on other nodes of $B$ via $f$ and the dependency graph of $B$ is $k$-regular is CPT-invariant. In particular, for all integers $k \geq 2$, the set of all $k$-$k$-XOR-BNs (Definition \ref{def:k-k-XOR-BNs}) and the set of all $(2k-1)$-$(2k-1)$-MAJORITY-BNs (Definition \ref{def:2k+1_2k+1_MAJORITY_BN}) are both CPT-invariant.
\end{example}

\begin{theorem}\label{thm:surjectivity_and_min_control}
Let $\Pi$ be a CPT-invariant set of BNs. For each BN $B$ in $\Pi$, let $V(B)$ be the set of nodes of $B$, $n(B)$ be the total number of nodes in $B$, $F^{B} : \{ 0, 1 \}^{n} \to \{ 0, 1 \}^{n}$ be the update rule of $B$, $U^{*}(B)$ be the smallest set of control nodes which can render $B$ controllable, and $W(B)$ be the largest set of nodes in $B$ such that the restriction function $F^{B}_{W(B)} : \{ 0, 1 \}^{n} \to \{ 0, 1 \}^{|W(B)|}$ is surjective. Define
$c \coloneqq \sup \left\{ \frac{|W(B)|}{n(B)} : B \in \Pi \right\}$. Then,
$\inf \left\{ \frac{|U^{*}(B)|}{n(B)} : B \in \Pi \right\} = 1 - c$. Thus, $|U^{*}(B)| \geq (1 - c)n(B)$ for all $B \in \Pi$.
\end{theorem}

\begin{proof}
Fix arbitrary BN $B \in \Pi$. By Theorem \ref{thm:restriction_becomes_surjective}, 
$F_{V(B) \setminus U^{*}(B)} : \{ 0, 1 \}^{n(B)} \to \{ 0, 1 \}^{n(B) - |U^{*}(B)|}$ is a surjective function. Then, we have
\begin{equation}
\frac{n(B) - |U^{*}(B)|}{n(B)} \leq c\ \Rightarrow\ \frac{|U^{*}(B)|}{n(B)} \geq 1 - c.
\end{equation}
Therefore, $1 - c$ is a lower bound of the set $\left\{ \frac{|U^{*}(B)|}{n(B)} : B \in \Pi \right\}$. Now, we will show that $1 - c$ is actually the greatest lower bound. Fix arbitrary $\varepsilon > 0$. Then, there exists a BN 
$\tilde{B} \in \Pi$ such that
\begin{align}
&c - \varepsilon < \frac{|W(\tilde{B})|}{n(\tilde{B})} \leq c \\
\Rightarrow\ &1 - c \leq \frac{n(\tilde{B}) - |W(\tilde{B})|}{n(\tilde{B})} < 1 - c + \varepsilon.
\end{align}
By Theorem \ref{thm:control_of_CSP_BNs}, for all integers $s \geq 2$, the BN $\text{CP}(\overbrace{F^{\tilde{B}}, F^{\tilde{B}}, \ldots, F^{\tilde{B}}}^{s\ \text{times}})$ is a BN in $\Pi$ with $sn(\tilde{B})$ nodes which admits a control-node set with size $n(\tilde{B}) + (n(\tilde{B}) - |W(\tilde{B})|)(s - 1)$. Then, we have
\begin{equation}
1 - c \leq \frac{|U^{*}(\tilde{B}_{s})|}{n(\tilde{B}_{s})} \leq
\frac{n(\tilde{B}) + (n(\tilde{B}) - |W(\tilde{B})|)(s - 1)}{sn(\tilde{B})}
= \frac{1}{s} + \left( 1 - \frac{1}{s} \right)\frac{(n(\tilde{B}) - |W(\tilde{B})|)}{n(\tilde{B})}.
\end{equation}
For $s$ sufficiently large, we have
\begin{equation}
\frac{1}{s} + \left( 1 - \frac{1}{s} \right)\frac{(n(\tilde{B}) - |W(\tilde{B})|)}{n(\tilde{B})} \leq
\varepsilon + \frac{(n(\tilde{B}) - |W(\tilde{B})|)}{n(\tilde{B})} < 1 - c + 2\varepsilon.
\end{equation}
Therefore,
\begin{equation}
1 - c \leq \frac{|U^{*}(\tilde{B}_{s})|}{n(\tilde{B}_{s})} < 1 - c + 2\varepsilon.
\end{equation}
Since $\varepsilon > 0$ is arbitrary, we deduce that $\inf \left\{ \frac{|U^{*}(B)|}{n(B)} : B \in \Pi \right\} = 1 - c$.
\end{proof}

\begin{remark}
Theorems \ref{thm:restriction_becomes_surjective} and \ref{thm:surjectivity_and_min_control} are useful because they allow us to derive general lower bounds for the minimum number of control nodes of BNs by studying the surjectivity of vector-valued Boolean functions. In Section \ref{section:MAJORITY}, we will use Theorem \ref{thm:restriction_becomes_surjective} to derive such a general lower bound for 
$(2k+1)$-$(2k+1)$-MAJORITY-BNs.
\end{remark}

\section{A New General Lower Bound for $(2k+1)$-$(2k+1)$-MAJORITY-BNs}\label{section:MAJORITY}

In this section, we will prove that there exists $d_{2k+1} \in (0, 1)$ such that for all controllable $(2k+1)$-$(2k+1)$-MAJORITY-BNs, the number of control nodes is at least $d_{2k+1}n$, where $n$ is the total number of nodes in the BN. We will first define the concept of $(2k+1)$-$(2k+1)$-MAJORITY-BNs (Definition \ref{def:2k+1_2k+1_MAJORITY_BN}). Then, we will prove some lemmas and corollaries (Lemmas \ref{lemma:share_one_input}, \ref{lemma:strictly_decreasing} and Corollary \ref{coroll:strictly_decreasing}) needed to prove the general lower bound. Afterwards, we will prove the key theorem (Theorem \ref{thm:m_over_n_<=_c}) needed for proving the general lower bound. Finally, we present the general lower bound result as a corollary (Corollary \ref{coroll:improved_lower_bound_MAJOR}) and compare it against existing general lower bounds derived for $(2k+1)$-$(2k+1)$-MAJORITY-BNs.

\begin{definition}\label{def:odd_majority_function}
For all integers $k \geq 1$, the $(2k+1)$-ary majority function $\theta_{2k+1} : \{ 0, 1 \}^{2k+1} \to \{ 0, 1 \}$ is defined to be the function such that $\theta_{2k+1}(\mathbf{x}) = 1$ iff 
$x_{1} + x_{2} + \cdots + x_{2k+1} \geq k + 1$.
\end{definition}

\begin{definition}\label{def:2k+1_2k+1_MAJORITY_BN}
If the constituent Boolean functions of a BN are all equal to $\theta_{2k+1}$ and the dependency graph of the BN is $(2k+1)$-regular, then the BN is said to be a $(2k+1)$-$(2k+1)$-MAJORITY-BN.
\end{definition}

\begin{lemma}\label{lemma:share_one_input}
Let $k \geq 1$ be an integer and $X_{1}$, $X_{2}$, \ldots, $X_{2k}$, $Y_{1}$, $Y_{2}$, \ldots, $Y_{2k}$, $A$ be mutually independent uniform binary random variables taking values in $\{ 0, 1 \}$. 
Then, 
\begin{equation}
\Pr(\theta_{2k+1}(X_{1}, X_{2}, \ldots, X_{2k}, A) = 0, \theta_{2k+1}(Y_{1}, Y_{2}, \ldots, Y_{2k}, A) = 1) < \frac{1}{4}.
\end{equation}
\end{lemma}

\begin{proof}
Let $D \coloneqq C^{2k}_{0} + C^{2k}_{1} + \cdots + C^{2k}_{k-1}$. Note that 
$D \coloneqq C^{2k}_{k+1} + C^{2k}_{k+2} + \cdots + C^{2k}_{2k}$ and hence
$2D + C^{2k}_{k} = 2^{2k}$. We have
\begin{equation}
D = \frac{2^{2k} - C^{2k}_{k}}{2}, \quad D + C^{2k}_{k} = \frac{2^{2k} + C^{2k}_{k}}{2}.
\end{equation}
Then, 
\begin{equation}
2D^{2} + 2D C^{2k}_{k} 
= 2D(D + C^{2k}_{k}) 
= \frac{(2^{2k} - C^{2k}_{k})(2^{2k}+C^{2k}_{k})}{2}
= 2^{4k-1} - \frac{(C^{2k}_{k})^{2}}{2}
< 2^{4k-1}.
\end{equation}
Hence,
\begin{align}
&\ 2D^{2} + 2^{2k}C^{2k}_{k} - (C^{2k}_{k})^{2}
= 2D^{2} + (2^{2k} - C^{2k}_{k})C^{2k}_{k} = 2D^{2} + 2DC^{2k}_{k} < 2^{4k}-1 \\
\Leftrightarrow
&\ 2D(D + C^{2k}_{k}) + 2^{2k}C^{2k}_{k} - 2DC^{2k}_{k} - (C^{2k}_{k})^{2} < 2^{4k-1} \\
\Leftrightarrow
&\ 2D(2^{2k} - D) + 2^{2k}C^{2k}_{k} - 2DC^{2k}_{k} - (C^{2k}_{k})^{2} < 2^{4k-1} \\
\Leftrightarrow
&\ \frac{2^{2k+1}D + 2^{2k}C^{2k}_{k} - 2D^{2} - 2DC^{2k}_{k} - (C^{2k}_{k})^{2}}{2^{4k+1}} < \frac{1}{4} \\
\Leftrightarrow
&\ \frac{(D + C^{2k}_{k})(2^{2k} - D - C^{2k}_{k}) + D(2^{2k} - D)}{2^{4k+1}} < \frac{1}{4} \\
\Leftrightarrow
&\ 
\Pr(A = 0) \Pr( X_{1} + X_{2} + \ldots + X_{2k} \leq k) \Pr(Y_{1} + Y_{2} + \ldots + Y_{2k} \geq k+1) + \nonumber \\
&\ \Pr(A = 1) \Pr( X_{1} + X_{2} + \ldots + X_{2k} \leq k-1) \Pr(Y_{1} + Y_{2} + \ldots + Y_{2k} \geq k) < \frac{1}{4} \\
\Leftrightarrow
&\ \Pr(\theta_{2k+1}(X_{1}, X_{2}, \ldots, X_{2k}, A) = 0, \theta_{2k+1}(Y_{1}, Y_{2}, \ldots, Y_{2k}, A) = 1) < \frac{1}{4}.
\end{align}
\end{proof}

\begin{lemma}\label{lemma:strictly_decreasing}
Let $k \geq 1$, $m \in [1, 2k-1]$ and $r = 2k-m$ be integers. Let $X_{1}, X_{2}, \ldots, X_{m}$, 
$Y_{1}, Y_{2}, \ldots, Y_{r}$, $Z_{1}, Z_{2}, \ldots, Z_{r}$, $A$, $B$ be mutually independent uniform binary random variables taking values in $\{ 0, 1 \}$. Set $\mathbf{X} \coloneqq (X_{1}, X_{2}, \ldots, X_{m})$,
$\mathbf{Y} \coloneqq (Y_{1}, Y_{2}, \ldots, Y_{r})$, $\mathbf{Z} \coloneqq (Z_{1}, Z_{2}, \ldots, Z_{r})$. Define
$u \coloneqq \Pr(\theta_{2k+1}(\mathbf{X}, A, \mathbf{Y}) = 1, \theta_{2k+1}(\mathbf{X}, B, \mathbf{Z}) = 1)$ and
$v \coloneqq \Pr(\theta_{2k+1}(\mathbf{X}, A, \mathbf{Y}) = 1, \theta_{2k+1}(\mathbf{X}, A, \mathbf{Z}) = 1)$.
Then, $v > u$.
\end{lemma}

\begin{proof}
Define $L_{1} \coloneqq \sum^{m}_{i = 1} X_{i} + \sum^{r}_{j=1} Y_{j}$ and 
$L_{2} \coloneqq \sum^{m}_{i = 1} X_{i} + \sum^{r}_{j=1} Z_{j}$. For all $c \in \{ 0, 1 \}$, we have
\begin{align}
\theta_{2k+1}(\mathbf{X}, c, \mathbf{Y})
&=
\begin{cases}
0	& \text{if } L_{1} \leq k-1, \\
c 	& \text{if } L_{1} = k, \\
1	& \text{if } L_{1} \geq k+1,
\end{cases} \\
\theta_{2k+1}(\mathbf{X}, c, \mathbf{Z})
&=
\begin{cases}
0	& \text{if } L_{2} \leq k-1, \\
c 	& \text{if } L_{2} = k, \\
1	& \text{if } L_{2} \geq k+1.
\end{cases}
\end{align}
Note that for all $a, b \in \{ 0, 1 \}$, $\Pr(A = a, B = b) = \frac{1}{4}$. Then, we have
\begin{align}
p_{1}	&\coloneqq \Pr(\theta_{2k+1}(\mathbf{X}, A, \mathbf{Y}) = 1 
\text{ and } \theta_{2k+1}(\mathbf{X}, B, \mathbf{Z}) = 1\ |\ L_{1} < k \text{ or } L_{2} < k) = 0, \\
p_{2}	&\coloneqq \Pr(\theta_{2k+1}(\mathbf{X}, A, \mathbf{Y}) = 1 
\text{ and } \theta_{2k+1}(\mathbf{X}, A, \mathbf{Z}) = 1\ |\ L_{1} < k \text{ or } L_{2} < k) = 0, \\
q_{1}	&\coloneqq \Pr(\theta_{2k+1}(\mathbf{X}, A, \mathbf{Y}) = 1 
\text{ and } \theta_{2k+1}(\mathbf{X}, B, \mathbf{Z}) = 1\ |\ L_{1} > k \text{ and } L_{2} > k) = 1, \\
q_{2}	&\coloneqq \Pr(\theta_{2k+1}(\mathbf{X}, A, \mathbf{Y}) = 1 
\text{ and } \theta_{2k+1}(\mathbf{X}, A, \mathbf{Z}) = 1\ |\ L_{1} > k \text{ and } L_{2} > k) = 1, \\
r_{1}	&\coloneqq \Pr(\theta_{2k+1}(\mathbf{X}, A, \mathbf{Y}) = 1 
\text{ and } \theta_{2k+1}(\mathbf{X}, B, \mathbf{Z}) = 1\ |\ L_{1} = k \text{ and } L_{2} > k) = \frac{1}{2}, \\
r_{2}	&\coloneqq \Pr(\theta_{2k+1}(\mathbf{X}, A, \mathbf{Y}) = 1 
\text{ and } \theta_{2k+1}(\mathbf{X}, A, \mathbf{Z}) = 1\ |\ L_{1} = k \text{ and } L_{2} > k) = \frac{1}{2}, \\
s_{1}	&\coloneqq \Pr(\theta_{2k+1}(\mathbf{X}, A, \mathbf{Y}) = 1 
\text{ and } \theta_{2k+1}(\mathbf{X}, B, \mathbf{Z}) = 1\ |\ L_{1} > k \text{ and } L_{2} = k) = \frac{1}{2}, \\
s_{2}	&\coloneqq \Pr(\theta_{2k+1}(\mathbf{X}, A, \mathbf{Y}) = 1 
\text{ and } \theta_{2k+1}(\mathbf{X}, A, \mathbf{Z}) = 1\ |\ L_{1} > k \text{ and } L_{2} = k) = \frac{1}{2}, \\
t_{1}	&\coloneqq \Pr(\theta_{2k+1}(\mathbf{X}, A, \mathbf{Y}) = 1 
\text{ and } \theta_{2k+1}(\mathbf{X}, B, \mathbf{Z}) = 1\ |\ L_{1} = k \text{ and } L_{2} = k) = \frac{1}{4}, \\
t_{2}	&\coloneqq \Pr(\theta_{2k+1}(\mathbf{X}, A, \mathbf{Y}) = 1 
\text{ and } \theta_{2k+1}(\mathbf{X}, A, \mathbf{Z}) = 1\ |\ L_{1} = k \text{ and } L_{2} = k) = \frac{1}{2}.
\end{align}
Therefore,
\begin{align}
u = 
&p_{1}\Pr(L_{1} < k \text{ or } L_{2} < k) + q_{1} \Pr(L_{1} > k \text{ and } L_{2} > k) + r_{1} \Pr(L_{1} = k \text{ and } L_{2} > k) + \nonumber \\
&s_{1} \Pr(L_{1} > k \text{ and } L_{2} = k) + t_{1} \Pr(L_{1} = k \text{ and } L_{2} = k) \nonumber \\
<
&p_{2}\Pr(L_{1} < k \text{ or } L_{2} < k) + q_{2} \Pr(L_{1} > k \text{ and } L_{2} > k) + r_{2} \Pr(L_{1} = k \text{ and } L_{2} > k) + \nonumber \\
&s_{2} \Pr(L_{1} > k \text{ and } L_{2} = k) + t_{2} \Pr(L_{1} = k \text{ and } L_{2} = k) \nonumber \\
=&v.
\end{align}
\end{proof}

\begin{corollary}\label{coroll:strictly_decreasing}
Consider an odd integer $2k+1$ where $k \geq 1$. For $m = 1, 2, \ldots, 2k$, let $X_{1}, X_{2}, \ldots, X_{m}$, 
$Y_{1}, Y_{2}, \ldots, Y_{2k+1-m}$, $Z_{1}, Z_{2}, \ldots, Z_{2k+1-m}$ be mutually independent uniform binary random variables taking values in $\{ 0, 1 \}$. Let $\mathbf{X} = (X_{1}, X_{2}, \ldots, X_{m})$, 
$\mathbf{Y} \coloneqq (Y_{1}, Y_{2}, \ldots, Y_{2k+1-m})$ and 
$\mathbf{Z} \coloneqq (Z_{1}, Z_{2}, \ldots, Z_{2k+1-m})$. For all $c, d \in \{ 0, 1 \}$, let 
$u^{m}_{c, d} \coloneqq \Pr(\theta_{2k+1}(\mathbf{X}, \mathbf{Y}) = c, \theta_{2k+1}(\mathbf{X}, \mathbf{Z}) = d)$.
Then, $u^{1}_{0, 1} > u^{2}_{0, 1} > \cdots > u^{2k}_{0, 1}$.
\end{corollary}

\begin{proof}
Since $\theta_{2k+1} : \{ 0, 1 \}^{2k+1} \to \{ 0, 1 \}$ is a self-dual Boolean function, $u^{m}_{0, 0} = u^{m}_{1, 1}$ and $u^{m}_{0, 1} = u^{m}_{1, 0}$. Hence, $u^{m}_{0, 1} = 0.5 - u^{m}_{1, 1}$ for $m = 1, 2, \ldots, 2k$.
By Lemma \ref{lemma:strictly_decreasing}, $u^{1}_{1,1} < u^{2}_{1,1} < \cdots < u^{2k}_{1,1}$. Therefore, we have
$u^{1}_{0, 1} > u^{2}_{0, 1} > \cdots > u^{2k}_{0, 1}$.
\end{proof}

\begin{definition}\label{def:incomplete}
Let $k \geq 1$ be an integer. If a function $F : \{ 0, 1 \}^{n} \to \{ 0, 1 \}^{m}$ is a function which satisfies the following three conditions
\begin{enumerate}
\item For all $x_{1}, x_{2}, \ldots, x_{n} \in \{ 0, 1 \}$,
\begin{multline}
F(x_{1}, x_{2}, \ldots, x_{n}) = \big(
\theta_{2k+1}(x_{j^{1}_{1}}, x_{j^{1}_{2}}, \ldots, x_{j^{1}_{2k+1}}), \\
\theta_{2k+1}(x_{j^{2}_{1}}, x_{j^{2}_{2}}, \ldots, x_{j^{2}_{2k+1}}),
\ldots, 
\theta_{2k+1}(x_{j^{m}_{1}}, x_{j^{m}_{2}}, \ldots, x_{j^{m}_{2k+1}})
\big);
\end{multline}

\item For all integers $i \in [1, m]$, $j^{i}_{1}, j^{i}_{2}, \ldots, j^{i}_{2k+1}$ are distinct;

\item For all integers $k \in [1, n]$, $k$ appears in the list $(j^{1}_{1}, j^{1}_{2}, \ldots, j^{1}_{2k+1}, j^{2}_{1}, j^{2}_{2}, \ldots, j^{2}_{2k+1}, \ldots, j^{m}_{1}, j^{m}_{2}, \ldots, j^{m}_{2k+1})$ no more than $2k+1$ times;
\end{enumerate}
then $F$ is said to be an incomplete $(2k+1)$-$(2k+1)$-MAJORITY function.
\end{definition}

\begin{theorem}\label{thm:m_over_n_<=_c}
There exists a real number $c \in (0, 1)$ such that for all surjective incomplete $(2k+1)$-$(2k+1)$-MAJORITY function $F : \{ 0, 1 \}^{n} \to \{ 0, 1 \}^{m}$, $\frac{m}{n} \leq c$.
\end{theorem}

\begin{proof}
Let $T_{i} \coloneqq \{ j^{i}_{1}, j^{i}_{2}, \ldots, j^{i}_{2k+1} \}$ where $i \in [1, m]$. Surjectivity of $F$ implies that $T_{1}, T_{2}, \ldots, T_{m}$ are distinct; otherwise, two output coordinates would always be equal.

Suppose that we have selected $k$ pairs of output indices $(p_{1}, q_{1}), (p_{2}, q_{2}), \ldots, (p_{k}, q_{k})$ such that $T_{p_{l}} \cap T_{q_{l}} \neq \varnothing$ and the sets $S_{l} = T_{p_{l}} \cup T_{q_{l}}$ are pairwise disjoint.
Let's sample the input uniformly from $\{ 0, 1 \}^{n}$, and consider the event
\begin{equation}
E = \bigcap^{k}_{l = 1} \{ Y_{p_{l}} = 0, Y_{q_{l}} = 1 \},
\end{equation}
where $Y_{j}$ is the random variable corresponding to the $j$-th output coordinate of $F$ for $j = 1, 2, \ldots, m$.
Since $1 \leq |T_{p_{l}} \cap T_{q_{l}}| \leq 2k$, by Lemma \ref{lemma:share_one_input} and Corollary \ref{coroll:strictly_decreasing}, we have
\begin{equation}
\Pr(Y_{p_{l}} = 0, Y_{q_{l}} = 1) \leq u^{1}_{0, 1} < \frac{1}{4},
\end{equation}
where $u^{1}_{0, 1}$ is defined in Corollary \ref{coroll:strictly_decreasing}. Because the sets $S_{l}$ are pairwise disjoint, these events are mutually independent. Consequently,
\begin{equation}\label{eq:majority_star_1}
\Pr(E) \leq (u^{1}_{0, 1})^{k}.
\end{equation}
On the other hand, exactly $2^{m - 2k}$ output vectors in $\{ 0, 1 \}^{m}$ satisfy the coordinate restrictions defining $E$. Surjectivity of $F$ gives each of these vectors at least one distinct pre-image, so
\begin{equation}\label{eq:majority_star_2}
\Pr(E) \geq \frac{2^{m-2k}}{2^{n}}.
\end{equation}
Combining Eq. (\ref{eq:majority_star_1}) and (\ref{eq:majority_star_2}) gives
\begin{align}
&\ 2^{m - 2k - n} \leq (u^{1}_{0, 1})^{k} \\
\Rightarrow&\ 2^{m-n} \leq (4u^{1}_{0, 1})^{k} \\
\Rightarrow&\ n - m \geq -\frac{k \ln(4u^{1}_{0, 1})}{\ln 2} \label{eq:majority_hashtag_1}
\end{align}
where $-\frac{k \ln(4u^{1}_{0, 1})}{\ln 2} > 0$ because $4u^{1}_{0, 1} < 1$.

Now, we will apply a greedy procedure. We start with $m$ sets $T_{1}, T_{2}, \ldots, T_{m}$. Whenever two remaining sets intersect, select them as a pair and delete every remaining set(s) that meets their union. Continue until the remaining sets are pairwise disjoint. Let $k$ be the number of selected pairs. For the $l$-th selection, 
let $s_{l} \coloneqq |S_{l}| \in [2k+2, 4k+1]$, and let $d_{l}$ be the number of sets deleted at that step, including the selected pair. By considering Property 3 in Definition \ref{def:incomplete}, we can see that
\begin{align}
&\ d_{l} - 2 \leq (2k+1)s_{l} - 2(2k+1) \\
\Rightarrow
&\ d_{l} \leq (2k+1)s_{l} - 4k \label{eq:majority_star_3}
\end{align}
Let $r$ be the number of sets remaining at the end of the greedy procedure. By construction, the sets $S_{l}$ are pairwise disjoint, and the remaining sets are disjoint both from one another and from all the $S_{l}$. Thus,
\begin{align}
m &= r + \sum^{k}_{l=1} d_{l}, \\
n &\geq (2k+1)r + \sum^{k}_{l = 1} s_{l}.
\end{align}
It follows that
\begin{align}
(2k+1)m - n
&\leq \sum^{k}_{l = 1} [(2k+1)d_{l} - s_{l}] \\
&\leq \sum^{k}_{l = 1} \left[ (2k+1)^{2} s_{l} - 4k(2k+1) - s_{l} \right]	& & \text{(by Eq. (\ref{eq:majority_star_3}))} \\
&= \sum^{k}_{l = 1} \left[ (4k^{2}+4k)s_{l} - 4k(2k+1) \right] \\
&\leq k\left[ (4k^{2}+4k)(4k+1) - 4k(2k+1) \right]				& & \text{(since $s_{l} \leq 4k+1$)} \\
&= 4k^{3}(4k+3). \label{eq:majority_hashtag_2}
\end{align}

Multiplying Eq. (\ref{eq:majority_hashtag_2}) by $-\frac{\ln(4u^{1}_{0, 1})}{\ln 2}$ and then using Eq. (\ref{eq:majority_hashtag_1}), we obtain
\begin{equation}
-\frac{\ln(4u^{1}_{0, 1})}{\ln 2}\left[ (2k+1)m - n \right]
\leq -\left( \frac{k \ln(4u^{1}_{0, 1})}{\ln 2} \right)[4k^{2}(4k+3)]
\leq 4k^{2}(4k+3)(n-m).
\end{equation}
Re-arranging, we get
\begin{equation}
\frac{m}{n} \leq \frac{4k^{2}(4k+3) - \frac{\ln(4u^{1}_{0, 1})}{\ln 2}}{4k^{2}(4k+3) - 
(2k+1)\frac{\ln(4u^{1}_{0, 1})}{\ln 2}} < 1.
\end{equation}
The proof is complete.
\end{proof}

\begin{corollary}\label{coroll:improved_lower_bound_MAJOR}
Let $2k+1 \geq 3$ and $n \geq 1$ be integers. Define
\begin{equation}
d_{2k+1} \coloneqq 1 - \frac{4k^{2}(4k+3) - \frac{\ln(4u^{1}_{0, 1})}{\ln 2}}{4k^{2}(4k+3) - 
(2k+1)\frac{\ln(4u^{1}_{0, 1})}{\ln 2}} = 
\frac{-2k\frac{\ln(4u^{1}_{0, 1})}{\ln 2}}{4k^{2}(4k+3) - (2k+1)\frac{\ln(4u^{1}_{0, 1})}{\ln 2}} \in (0, 1).
\end{equation}
For all $n$-node controllable $(2k+1)$-$(2k+1)$-MAJORITY-BNs with control-node set $U$,
$|U| \geq d_{2k+1}n$.
\end{corollary}

\begin{proof}
This corollary follows directly from Theorems \ref{thm:restriction_becomes_surjective} and \ref{thm:m_over_n_<=_c}.
\end{proof}

Our general lower bound for $(2k+1)$-$(2k+1)$-MAJORITY-BNs (Corollary \ref{coroll:improved_lower_bound_MAJOR}) is different from the following old general lower bound derived in
\cite{FokMinControl} because our bound does not require that any state of the BN can reach any other state under the influence of control signals within a bounded time period. The old general lower bound is given below.

\begin{theorem}
Let $n, k, s$ be integers such that $s \geq 1$ and $n \geq 2k+1 \geq 3$. Consider any $n$-node controllable 
$(2k+1)$-$(2k+1)$-MAJORITY-BN such that for all
$\mathbf{a} \neq \mathbf{b} \in \{ 0, 1 \}^{n}$, there exist control signals 
$\mathbf{u}(0)$, $\mathbf{u}(1)$, \ldots, $\mathbf{u}(t)$ (where $t \leq s$) that drive the BN from $\mathbf{a}$ (time $0$) to $\mathbf{b}$ (time $t+1$). Then, the control-node set $U$ satisfies
$|U| \geq \max(k+1, \left\lceil \frac{n+1}{s+1} \right\rceil)$.
\end{theorem}

\section{Conclusions}\label{section:conclusions}

Boolean networks (BNs) provide an important modeling framework for gene regulatory networks and a variety of other biological systems. This paper addresses the problem of determining the minimum number of nodes needed to control BNs under degree constraints. Our contributions are threefold. For XOR-BNs with $k$-regular dependency graphs, we obtain three new best-case upper bounds. We also introduce the notion of invariance under cyclically partite transformation (CPT-invariance) for sets of BNs and establish links between the controllability of BNs and the surjectivity of vector-valued Boolean functions. Using these surjectivity considerations together with a probabilistic argument, we further establish a new general lower bound on the minimum number of control nodes for majority-type threshold BNs whose dependency graphs are $(2k+1)$-regular. We expect our research to provide some contributions towards minimal interventions in real-world phenomenon such as biological systems and opinion dynamics in social networks.

There are at least two possible future directions of research related to this paper. Firstly, the research community can further explore whether considering the surjectivity of vector-valued Boolean functions is useful for deriving further results related to the minimal controllability of BNs. Secondly, for $k \geq 5$, further research can be conducted on the minimal controllability of XOR-BNs with $k$-regular dependency graphs.

\end{document}